\documentclass[prl,twocolumn,superscriptaddress,showpacs, nofootinbib]{revtex4-2}
\usepackage{graphicx}
\usepackage[utf8]{inputenc}
\usepackage{dcolumn}
\usepackage{hyperref}
\usepackage[american]{babel}
\usepackage{booktabs}
\hypersetup{
    colorlinks=true,
    linkcolor=blue,
    filecolor=magenta,      
    urlcolor=cyan,
    citecolor=magenta,
}
\usepackage{xcolor}
\usepackage{soul}
\usepackage{amsthm,amssymb}
\providecommand{\U}[1]{\protect\rule{.1in}{.1in}}

\usepackage{epigraph}
\begin{document}
\epigraph{``Nullum est oleae fructu utilius''\footnote{
``The olive is the richest gift of heaven.''}\\
	Pliny the Elder, Natural History, Book XV}

\title{The Physics of Olive Oil}
\author{Sergey Parnovsky}
\email{parnovsky@gmail.com}
\affiliation{Department of Astrophysics, National Taras Shevchenko University of Kyiv, Observatorna St. 3, Kyiv, 04053, Ukraine}
\author{A. A. Varlamov}
\email{andrey.varlamov@spin.cnr.it}
\affiliation{CNR-SPIN, Via del Fosso del Cavaliere, 100, 00133 Rome, Italy}
\affiliation{Istituto Lombardo “Accademia di Scienze e Lettere”, Via Borgonuovo 25, Milan, Italy}

\date{\today}
\begin{abstract} 
Olive oil is an integral part of Mediterranean culture, shaped by more than three thousand years of history, rich traditions, evolving technologies, and fundamental physical principles. This article explores the role of physics in the production of olive oil, highlighting how physical laws govern each stage of the process.
\end{abstract}

\maketitle

Since the time Greek colonists brought olive trees to the Italian peninsula, their cultivation and the extraction of the magical green-golden liquid—olive oil—have become an integral part of Mediterranean life. Olive trees symbolized peace, prosperity, and divine grace, appearing in temples and rituals, in the olive wreaths that crowned athletes (a Greek tradition continued in Rome), and even in funerary rites, where olive oil was used for anointing.

Roman patricians took pride in their olive groves \footnote{Cato the Elder (234–149 BCE), already in the second century before Christ, in his work ``De Agri Cultura'' (On Agriculture) gave a manual for Roman farmers: ``The olive orchard should be in a warm and sunny place, and the land should be good for it... the best oil is from ripe olives, not overripe''. 
	Marcus Terentius Varro (116–27 BCE) in ``Rerum Rusticarum Libri Tres'' (Three Books on Agriculture) discussed the importance of olives to Roman farming estates and highlighted the economic value of olives for landowners.
} , and this tradition continued through the Middle Ages, from the Republic of Genoa to the Kingdom of Sicily. Even today, at the end of October, university professors and bank clerks, engineers and medical doctors return from cities like Rome, Naples, and Milan to their family estates to harvest olives. Together with wine and bread, olive oil still forms one of the three pillars of the Mediterranean diet.
Although the chemical composition of modern extra virgin olive oil is almost identical to that produced centuries ago, the extraction process has changed fundamentally. Today, every step is  mechanized and computer-controlled. The process is faster, more reliable, and—most importantly—prevents oxidation by avoiding contact with air. 
\begin{figure}
	\includegraphics[width=0.99\columnwidth]{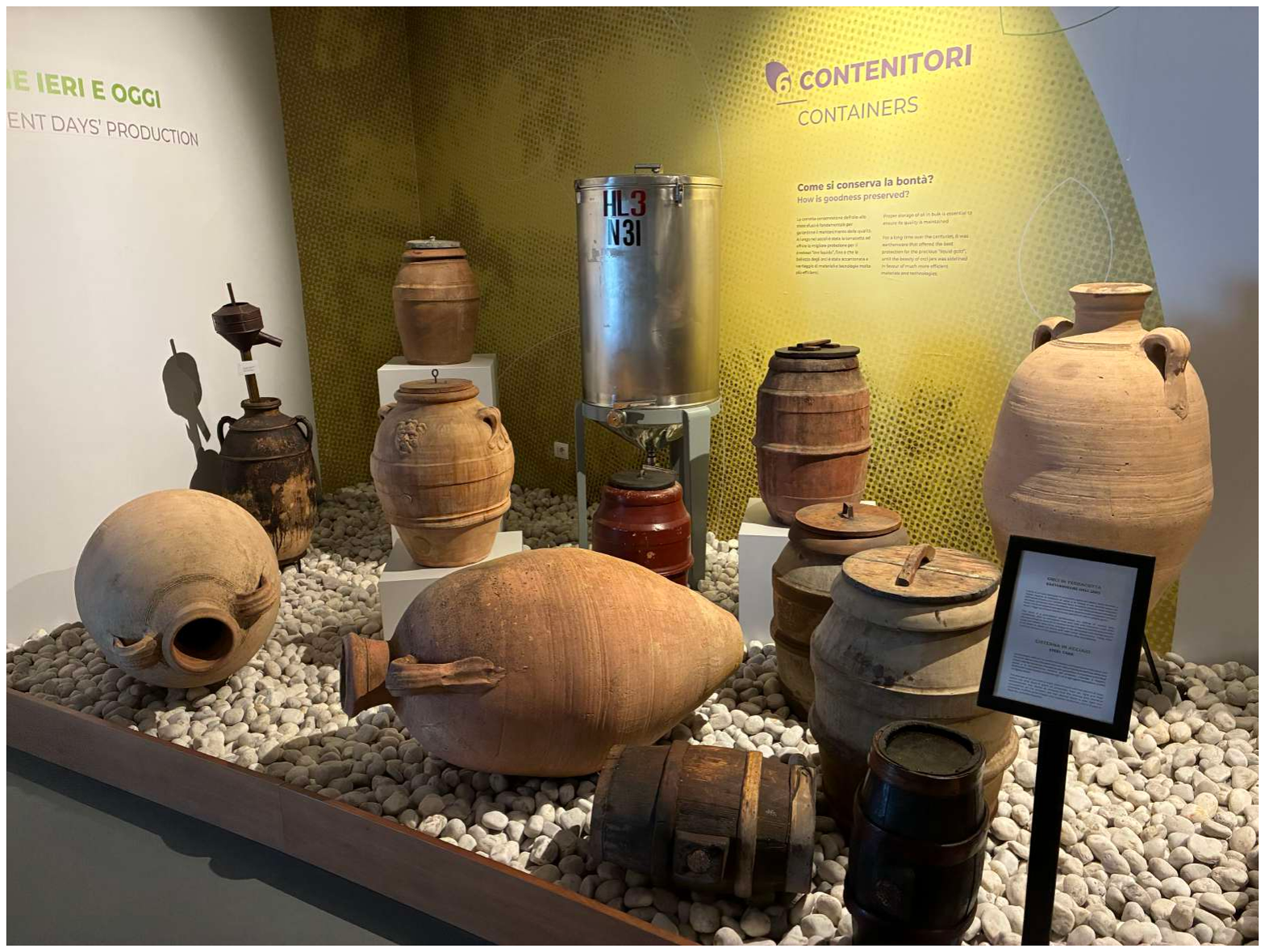}
	\caption {The Museum of Olive Oil in Monte San Savino.}
\label{museum}
\end{figure}
	
\section{The oil extraction process as technology}

The evolution of olive oil production can be vividly explored in the Olive Oil Museum of Monte San Savino, a small Tuscan town where both traditional and modern methods are displayed.

Oil extraction process required a series of separate steps performed historically on different pieces of equipment, often with long pauses in between. In modern facilities, by contrast, the entire process is continuous and sealed from air exposure. The olives that enter the production line gradually transform into high-quality extra virgin olive oil.

It all begins with the harvesting of high-quality olives, but we won't cover that part; our tour begins outside the oleificio (oil mill) gates. The pressing process can be roughly divided into four groups of technological operations.

\begin{figure}
	\begin{center}
		\includegraphics[width=0.99\columnwidth]{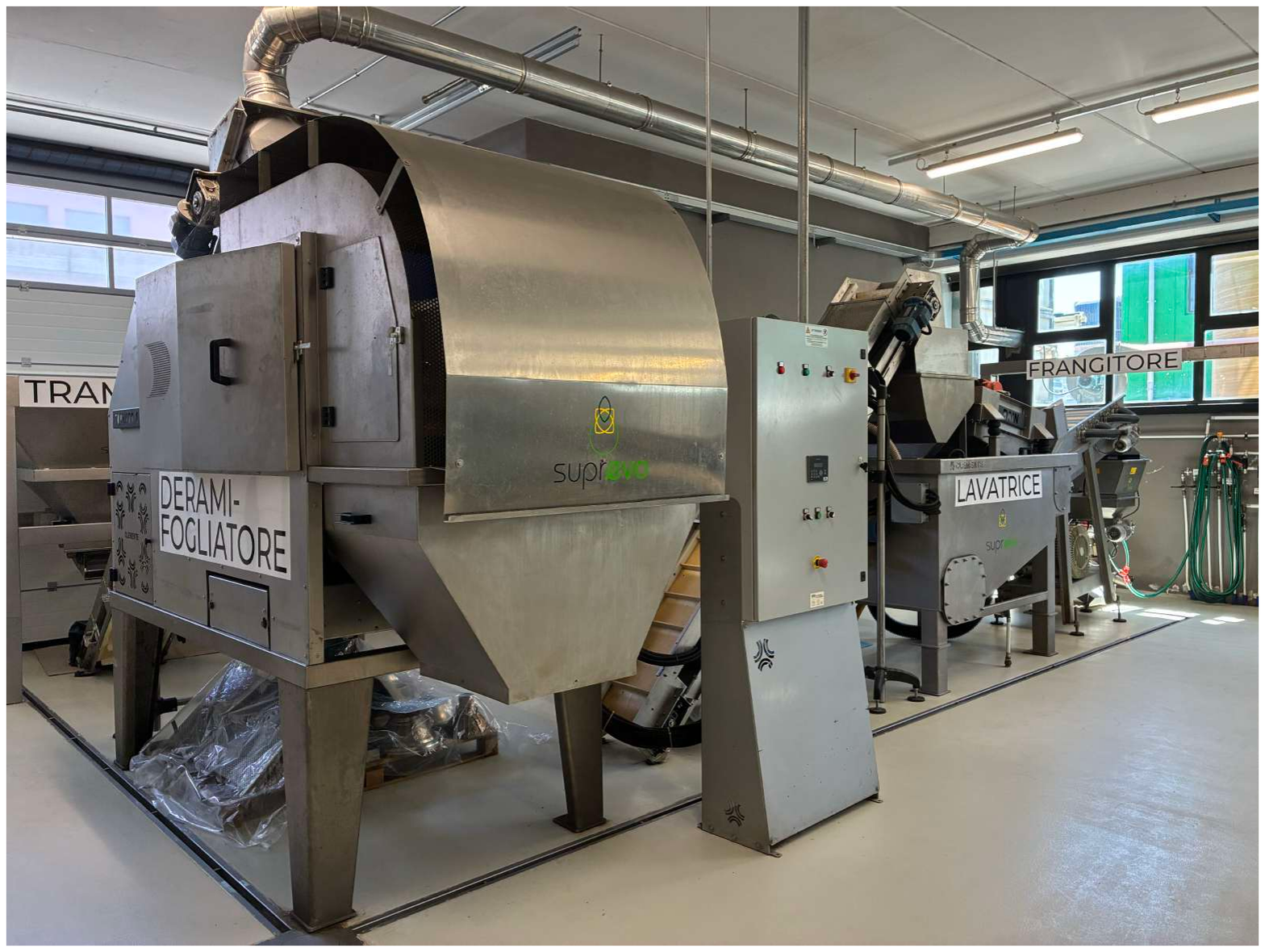}
	\end{center}
	\caption{Washing machine.}
	\label{washing}
\end{figure}

\begin{itemize}
\item {The first stage is the preparation and initial crushing of the olives. They are separated from any leaves and twigs that may have been accidentally collected during harvesting, washed, and crushed (see Figure 2). In the old days, the olives were not washed, but now this step has been added to the traditional process. Everything is done automatically. This part of the process is not particularly interesting from a physics point of view.}

\item{The next stage is the homogenization of the crushed olives into a homogeneous paste  (which is called sansa or pulp). What does it consist of? Naturally, it contains olive oil. After all, everything is done for the sake of extracting it. 100 kg of olives contain 12-18 kilograms of oil, depending on the variety, place of harvest, and weather conditions during the year of harvest. But there is much more water in the olives – from 35\% to 50\% of the total mass. The rest consists of solid residues of the peel, pits, and pulp, which are called "pomace" or "oil meal". After separation from water and oil, the solid and moist pomace, with a small amount of residual oil, is used for further processing. Biofuel (e.g., for fireplaces) is obtained from the crushed pit residues contained in it (Figure 3), and the remaining oil that could not be separated at the main stage of the pressing process is extracted from the soft fraction in one way or another \footnote{Extracting second-quality olive oil (also known as lampante oil or residual oil) from the residue (pomace) left after the first pressing was a known practice in ancient and modern methods alike. After the first cold press, pomace was pressed again, often heated slightly or mixed with hot water to loosen remaining oil. This second pressing produced lower-quality oil used for: lamps (hence lampante oil), soap, and for cooking by poorer households. Romans also would boil the pomace with water to help release the trapped oil. Once cooled, the oil floated on top and could be skimmed off. This method increased yield but further degraded the oil’s flavor and shelf-life. Pliny the Elder distinguishes oil qualities: "the first oil is for food, the rest for lamps". Nowadays exists also chemical extraction: use of solvents like hexane to extract every last drop of oil from the dry pomace, what results in refined pomace oil}. Naturally, this is no longer extra virgin oil, but a less valuable product.

\begin{figure}
	\begin{center}
		\includegraphics[width=0.90\columnwidth]{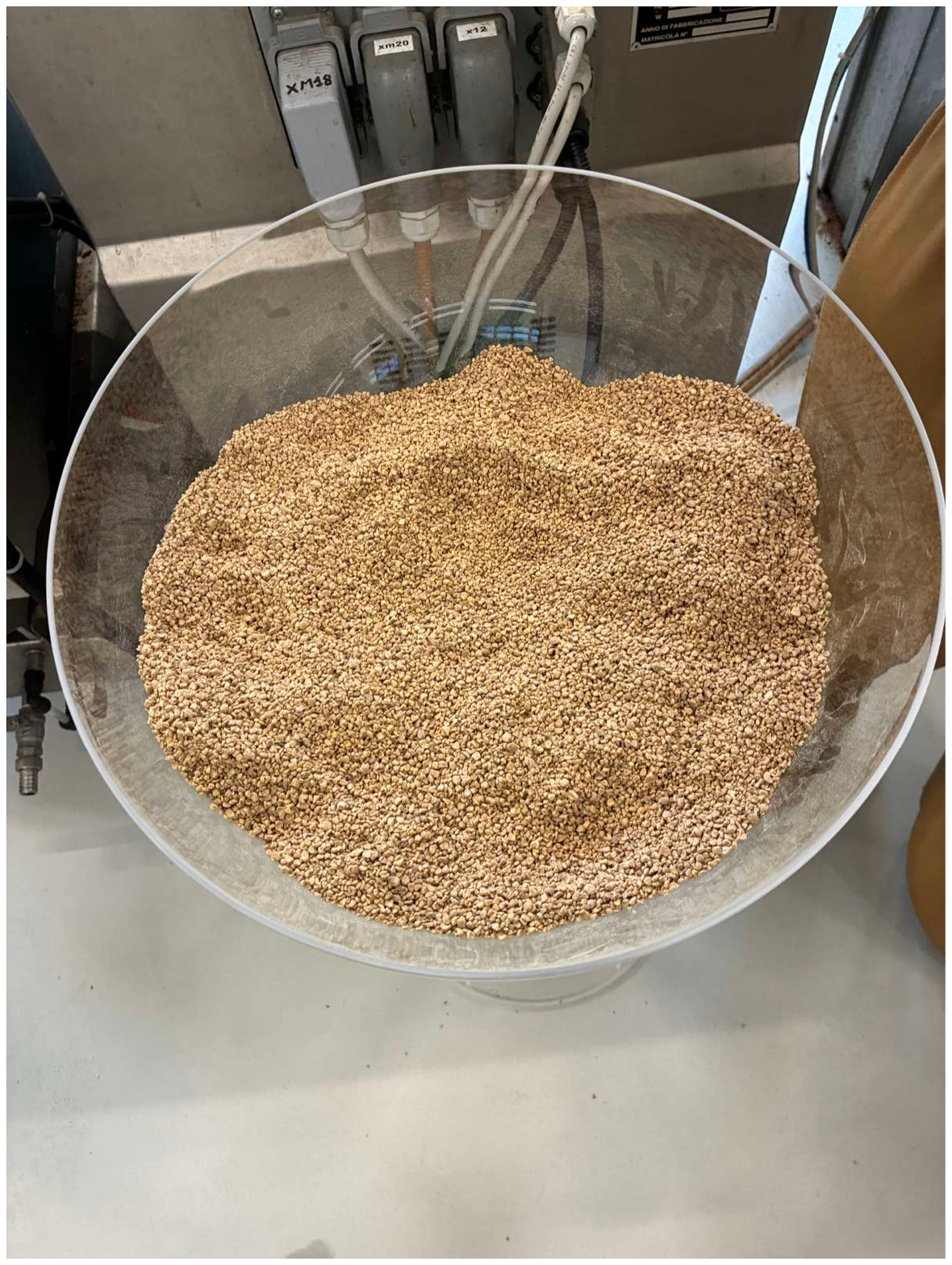}
	\end{center}
	\caption{Biofuel.}
	\label{Biofuel}
\end{figure}

In a traditional Italian frantoio (artisan oil mill), the crushing was done with heavy stone wheels. At the Monte San Savino Museum, an ancient millstone 110 cm in diameter, 33 cm wide, and weighing about 750 kg can still be seen (see Figure 4). Driven by animals, it slowly crushed olives and pits into a paste.

In modern oleificio rotating knives spin at about 4000 rpm. In 15 minutes, they produce a homogeneous paste while simultaneously promoting the coalescence of tiny oil droplets into larger ones—an important preparation for subsequent separation. Instead of hammer mills, which were used previously, knives are now mainly used (see Figure 5), which cause less damage to the pits and skins, thus enhancing the oil’s chlorophyll content and giving it a greener hue.}

\item{The third stage is the partitioning of the paste (sansa). Previously, this was done using a mechanical or hydraulic press (see Figure 6). The paste was first placed in special bags made of palm or hemp fiber, which were then placed under the press. The result of this pressing was a suspension of oil, water, and small residues of pomace, while most of the pomace remained trapped in the fibers of the cloth.}

\item{The fourth step consists of extracting the oil from the resulting suspension and filtering it. To this end, the difference in density between olive oil (which varies slightly depending on the characteristics of the olives, but on average is close to 0.91\ g/cm³) and water, whose density is 1\ g/cm³, was traditionally exploited. In the past, the suspension was left to settle in large containers, such as vats or terracotta jugs, for hours or even days. During this time, a natural separation occurred: the oil rose to the top, while the water and sediment settled to the bottom. All that remained was to pour or "scoop" the oil with ladles, without touching the water layer. Sometimes the decanting process was repeated several times to obtain a purer oil. In any case, it was impossible to avoid contact between the oil and air.}

\end{itemize}

In modern oil mills, the oil extraction process is performed differently and much more rapidly. The division into phases has also changed. Traditional decantation has been divided into preliminary and final oil extraction, which are performed with different equipment. The preliminary extraction of oil from the pulp is combined with its separation from the water. In the Monte San Savino Museum, you can see the modern machines used in all stages of olive oil transformation. We will briefly describe their operation for our readers. 

After exiting the crushing machine, the pulp enters a device that is still called a decanter (see Figure 7). But behind the old name lies completely different processes. Today, a decanter is a horizontal centrifugal separator for separating pulp into solid residues, water, and olive oil. The pulp is fed under pressure into the front of the horizontal drum, which rotates at an enormous angular velocity of $3,000–3,500 rpm$. Consequently, particles not in the immediate vicinity of the axis are subjected to centrifugal acceleration of $1,000–4,000 g$ (where the letter $g=9.81 m/s^2$ indicates the acceleration due to gravity). Under the influence of the resulting centrifugal forces, the heavier components (solid residues and water) move toward the wall of the drum, while the lighter first-press oil (still a suspension containing tiny impurities and mucous substances) moves closer to the axis of rotation. This suspension is also subjected to centrifugal forces that attempt to push it away from the axis, but the Archimedean forces generated during floating in water are stronger and push it toward the center.

\begin{figure}
\begin{center}
\includegraphics[width=0.95\columnwidth]{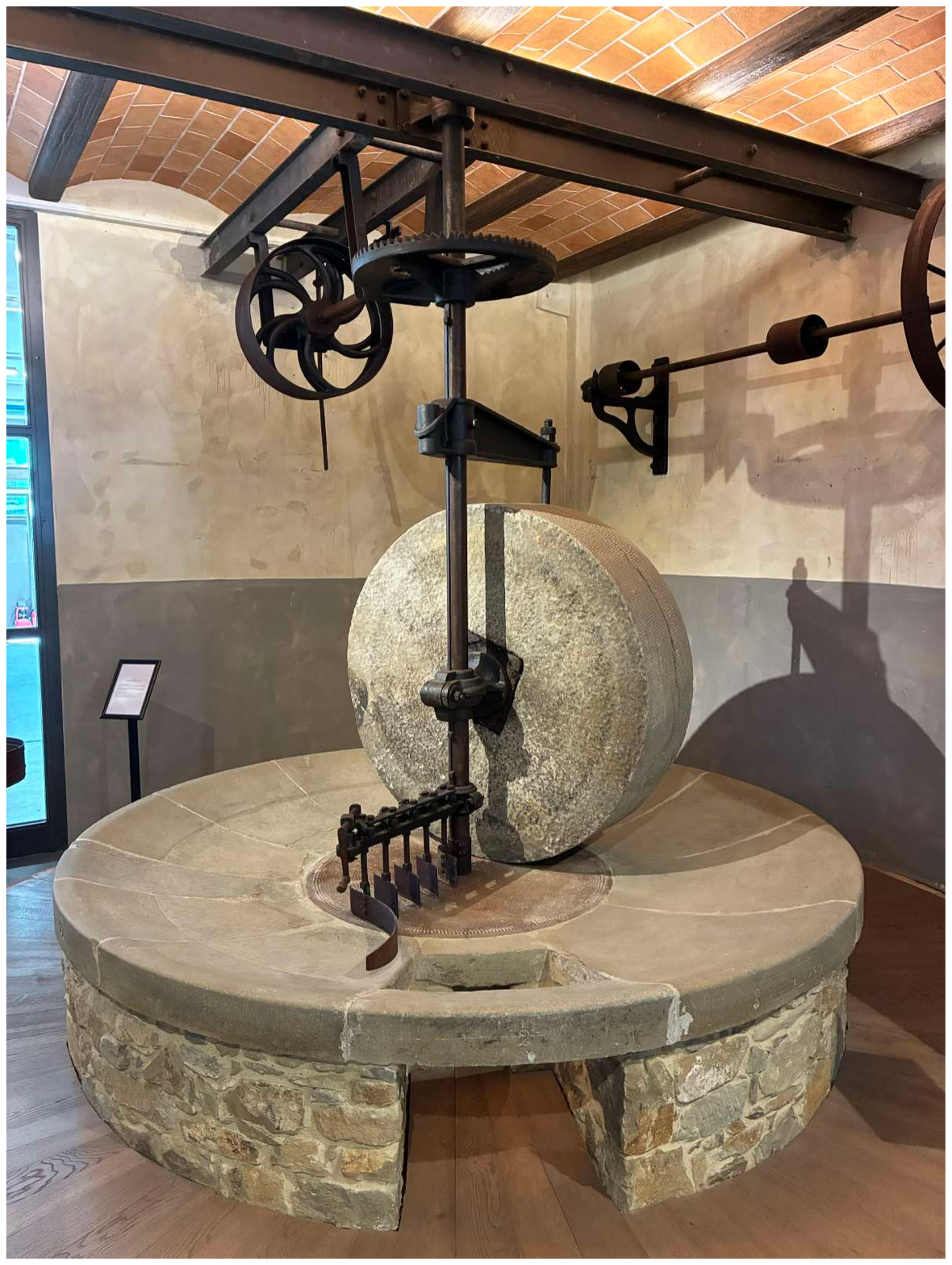}
\end{center}
\caption{The stone wheel for olive crushing.}
\label{stone}
\end{figure}

Inside the drum, a screw conveyor rotates at a slightly different angular speed (see Figure 8). This collects and pushes the solid particles toward the narrow conical section, where they exit. 
\begin{figure}
	\begin{center}
		\includegraphics[width=0.95\columnwidth]{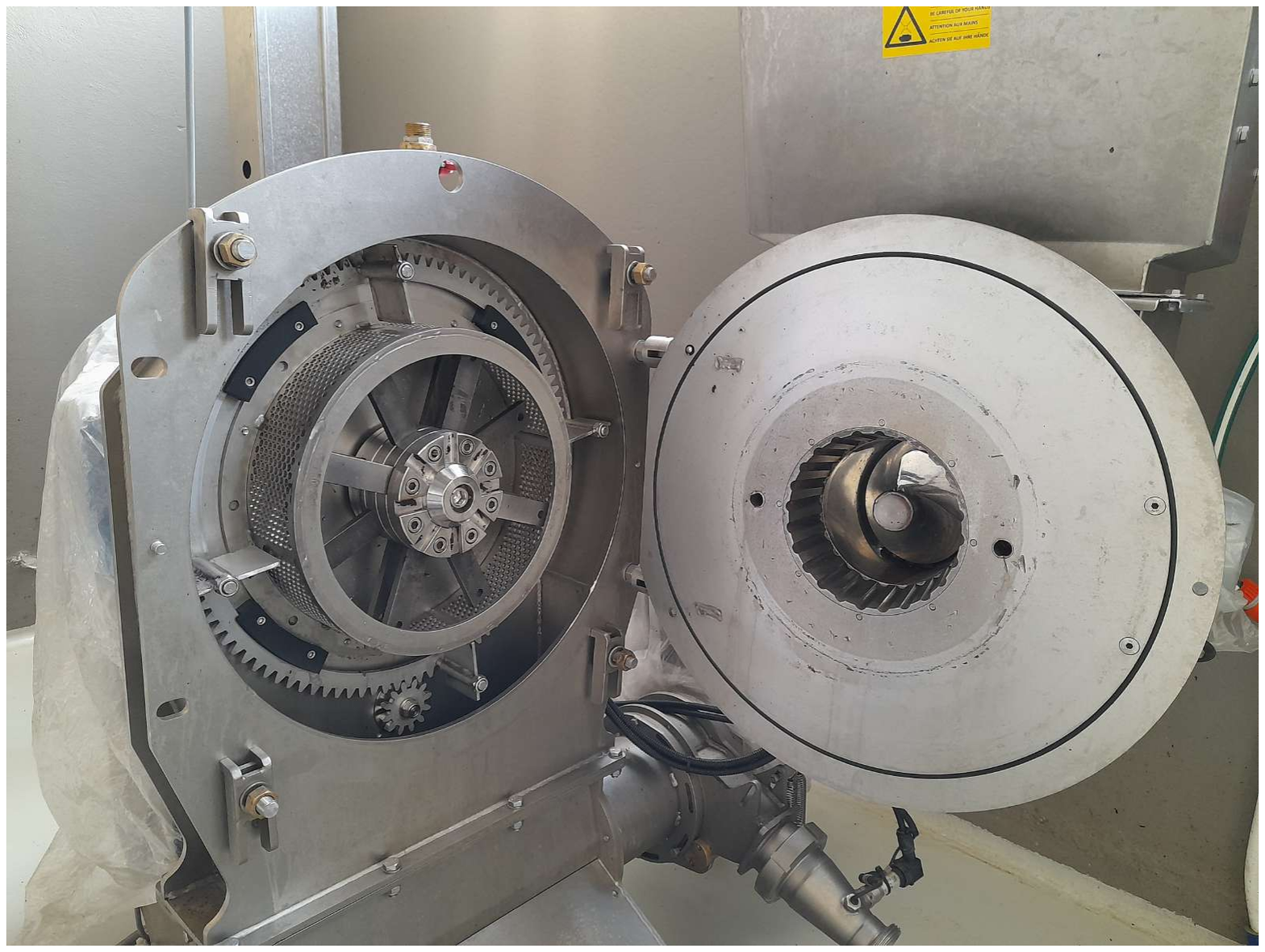}
	\end{center}
	\caption{The modern  slow crusher (frangitore).}
	\label{frangitore}
\end{figure}

The suspension then enters a second, slowly rotating chamber, designed for more thorough cleaning of the oil. As the suspension passes through this chamber, even the smallest impurities and mucus are removed. Upon exiting this device, the liquid is filtered to remove residual solid particles and then passes into a vertical centrifuge, where residual water is removed by centrifugal forces (see Figure 9). Filtration and centrifugation can be considered the final stage of the modern oil extraction process.

\begin{figure}
	\begin{center}
		\includegraphics[width=0.95\columnwidth]{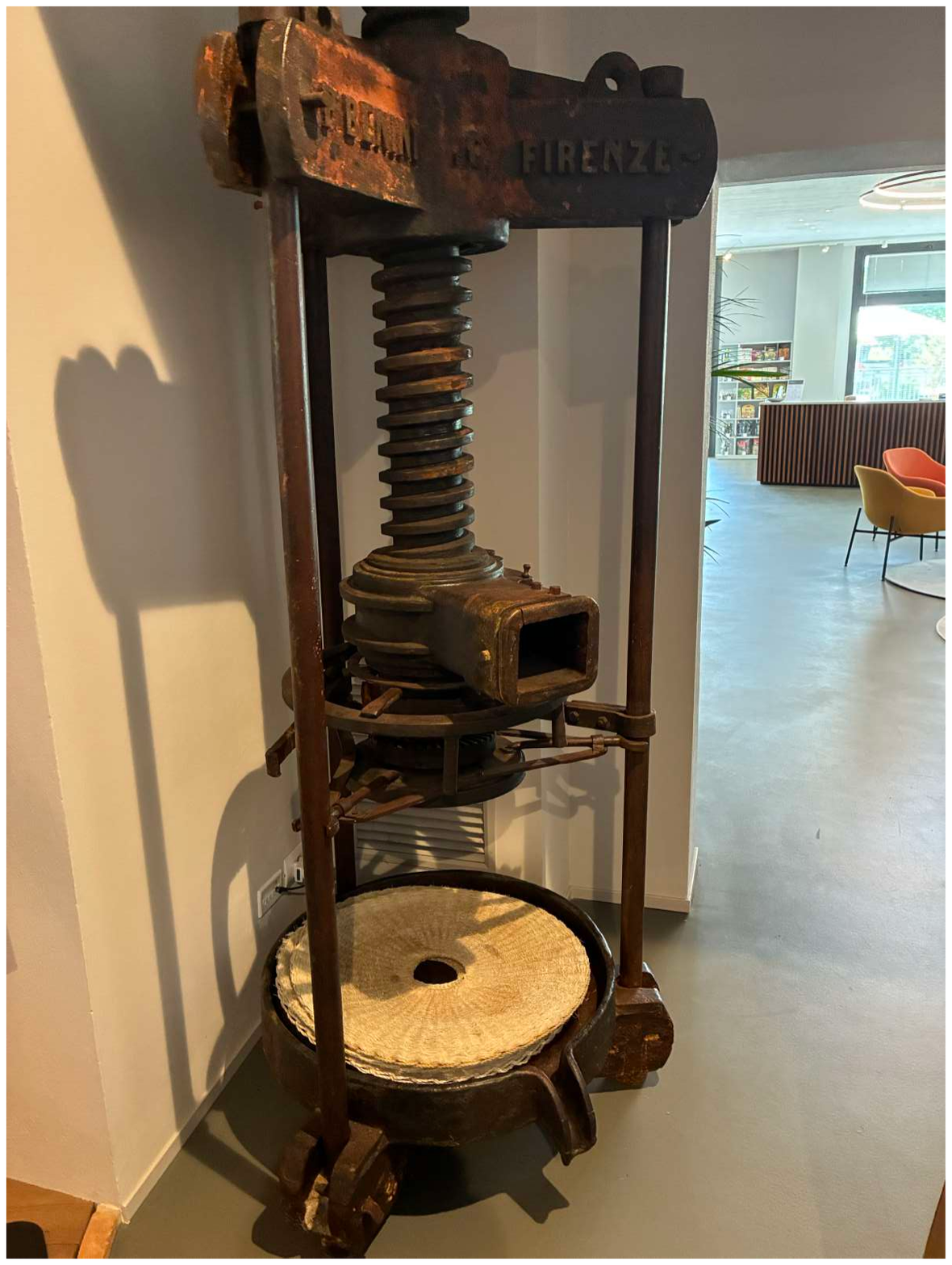}
	\end{center}
	\caption{Ancient mechanical press.}
	\label{press}
\end{figure}

Before bottling (in a nitrogen atmosphere to prevent contact with oxygen), the quality of the resulting oil is verified both with chemical analysis and good old-fashioned organoleptic methods. Simply put, professionals evaluate every nuance of its flavor and aroma. 

As we can see, modern methods allow us to extract oil from olives more quickly, with greater productivity and avoiding its oxidation during the process.

\begin{figure}
	\begin{center}
		\includegraphics[width=0.95\columnwidth]{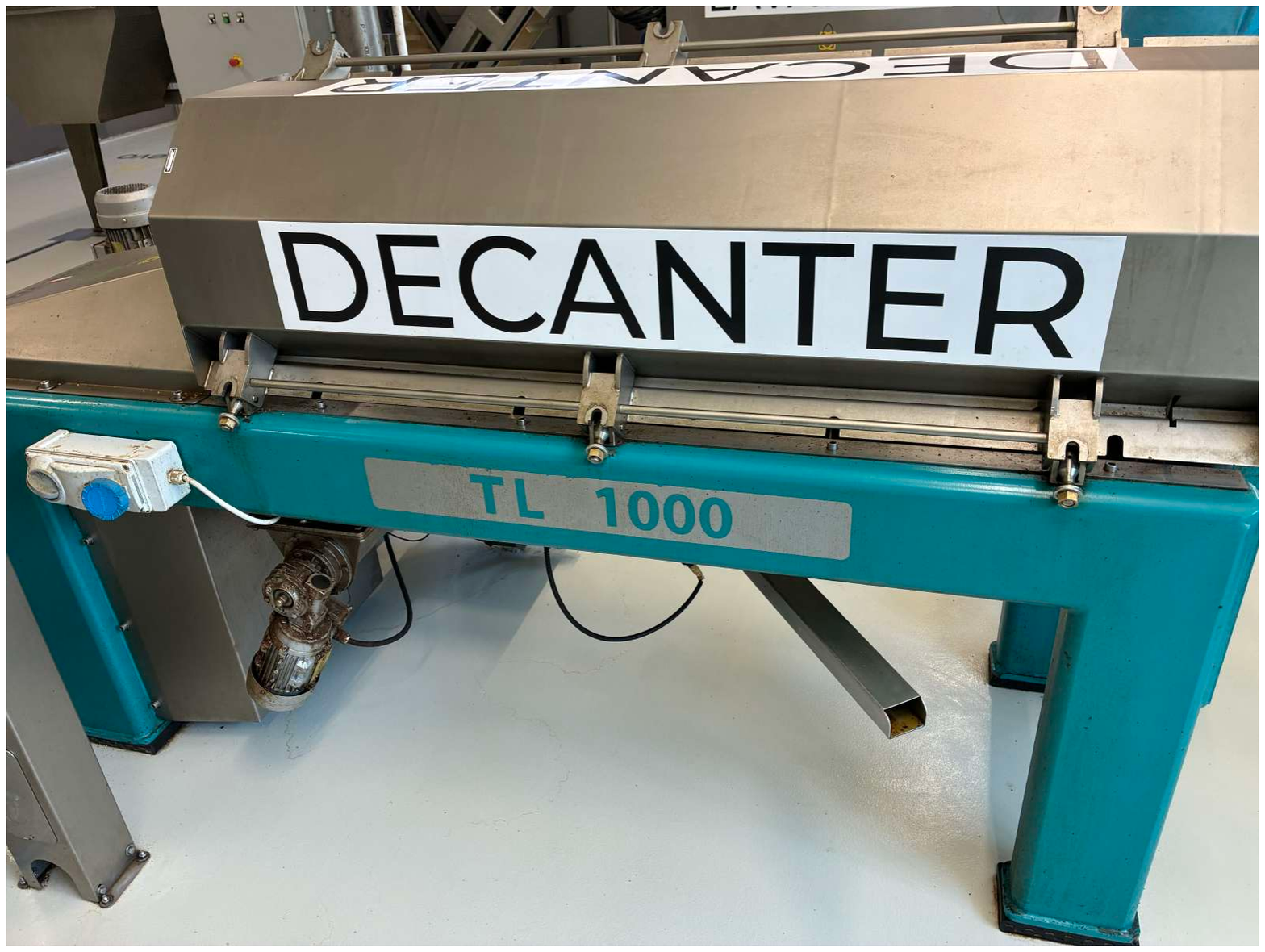}
	\end{center}
	\caption{Modern decanter TL 1000.}
	\label{decanter}

\end{figure}
\begin{figure}
	\begin{center}
		\includegraphics[width=0.95\columnwidth]{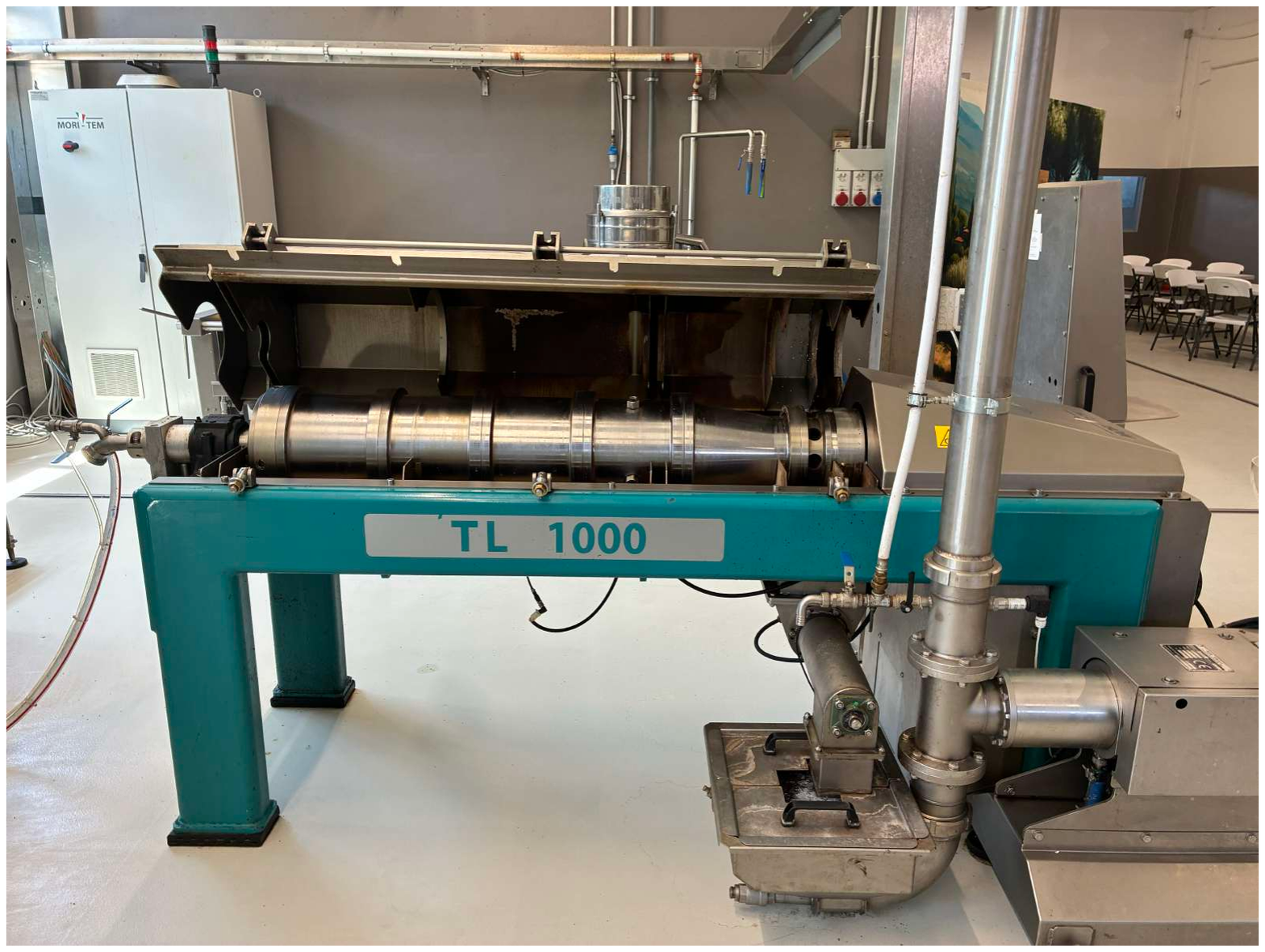}
	\end{center}
	\caption{A screw conveyor inside the drum.}
	\label{screw}
\end{figure}

\begin{figure}
\begin{center}
	\includegraphics[width=0.95\columnwidth]{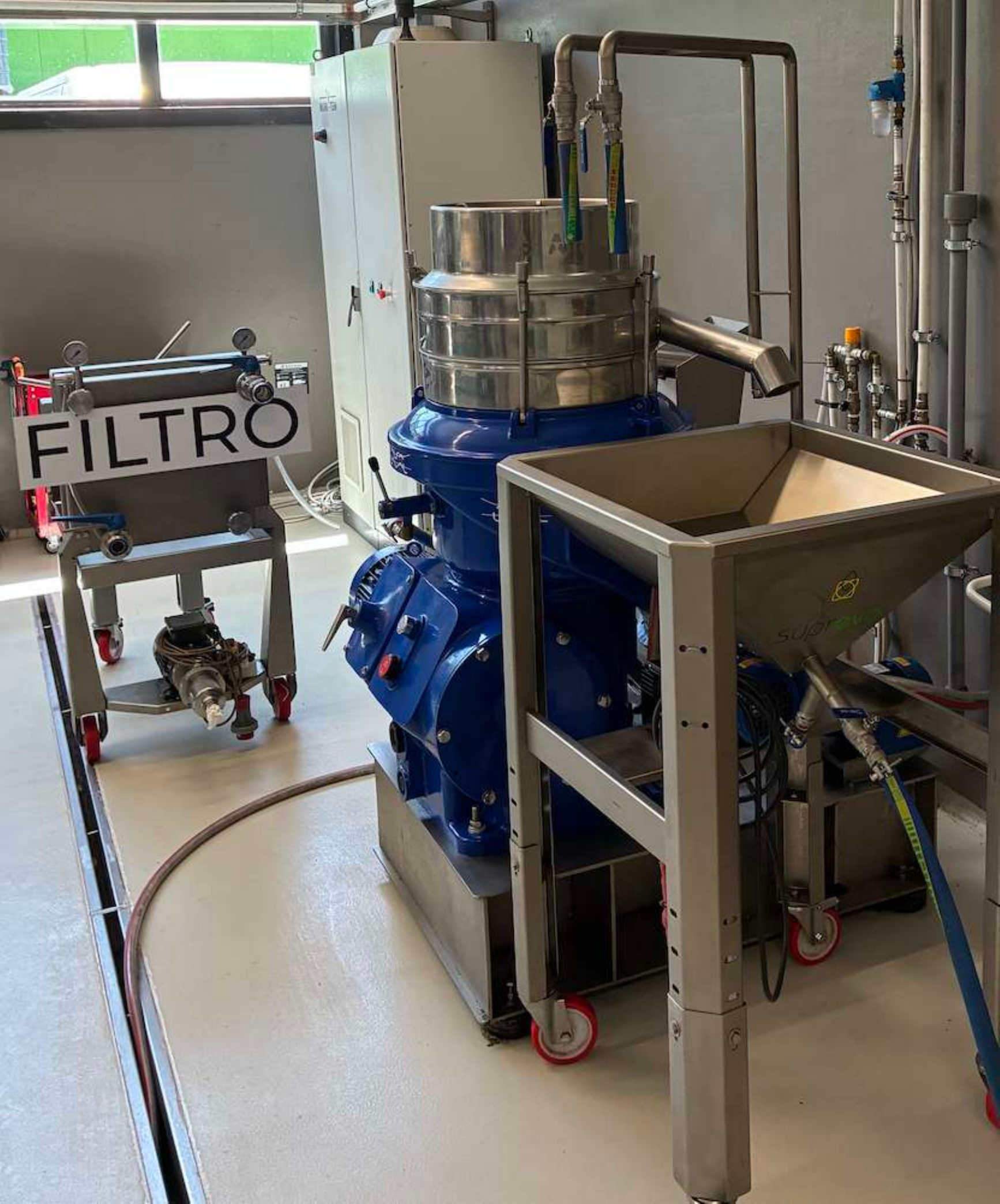}
\end{center}
\caption{The separator – vertical centrifuge.}
\label{separator}
\end{figure}

\section{Oil extraction as a series of physical processes}

Let's examine the process described from a physics perspective. We began with the decantation of the suspension of oil, water, and pomace residue. Why did it take so long? Let's imagine this process in its advanced stages. The separation is almost complete, and we see a layer of oil atop a layer of liquid composed primarily of water. In the water, there are still oil droplets floating slowly, while in the oil, the water droplets slowly sink. This picture is vaguely reminiscent of the gas bubbles floating in sparkling wine or carbonated water. 

However, the movement of both the water and oil droplets will be noticeably slower. The small gas bubbles in the sparkling wine have an almost spherical shape and rise because they are pushed by the Archimedean force, which greatly exceeds the insignificant weight of the gas inside the bubble. Drops of olive oil float in water for the same reason. The average density of oil is $0.91 \ g/cm^3$, less than the density of water, which is $1\ g/cm^3$, although only by 9\%. Conversely, a heavier drop of water in olive oil will sink. The difference between the Archimedean force acting on the drop and its weight is called buoyancy. If the buoyancy is positive, the drop floats; if it is negative, it sinks.

The buoyancy that sets the drop in motion shortly after its inception, when the drop's velocity stabilizes, is balanced by the drag force. Suppose that in the problem at hand, a spherical drop (or a bubble) with density $\rho_0$, radius $r$, and viscosity coefficient $\eta_0$, moves in a medium with density $\rho_1$ and viscosity coefficient $\eta_1$. In this case, it is not easy to find the drag force. However, back in 1911, for not too high velocities, it was calculated independently of each other by the Frenchman Hadamard \cite{Hadamard}  and the Pole Rybchynski \cite{Rybchynski} . Consequently, it is possible to use their result and obtain the following formula for the drop's buoyancy speed (with $\rho_0>\rho_1$) or its fall (with $\rho_0<\rho_1$):
\begin{equation}
v=\frac{2}{3}\frac{r^2g\left(\rho_0-\rho_1\right)}{\eta_0}\frac{\eta_0+\eta_1}{2\eta_0+{3\eta}_1}.
\end{equation}   

Note that the viscosity of the droplet $\eta_1$ only affects the last multiplier and not significantly. If it is considerably higher than the viscosity of the medium $\eta_0$, this coefficient is close to $1/3$; if it is considerably lower than the latter, then it is close to $1/2$. In any case, the values of this multiplier are between $1/3$ and $1/2$. More important is the dependence on the ratio of the densities and the viscosity of the medium \footnote{We note that for a rigid sphere ($\eta_1\rightarrow\infty$) this complex formula naturally transforms into the well-known Stokes formula ($F=6 \pi \eta_0 r v$):

\begin{eqnarray}
v=\frac{2}{9}\frac{r^2g\left(\rho_0-\rho_1\right)}{\eta_0}=\frac{1}{6\pi\eta_0r} \nonumber \ \ \frac{4\pi}{3}r^3g\left(\rho_0-\rho_1\right)=\frac{(F_A-mg)}{6\pi\eta_0 r}.
\nonumber
\end{eqnarray}}. 

Let's take, for example, a small bubble or droplet with a diameter of 0.2\ mm, or r\ =\ 0.1\ mm. Its shape is practically spherical, and the speed of movement is so low that we can apply the formula given above. When the bubble floats, the density and viscosity of the gas are negligible compared to the corresponding characteristics of water. For the buoyancy velocity of the gas bubble, the calculation gives a value close to 3.3\ cm/s. An oil droplet floats in water 16.5 times slower due to its 11 times lower buoyancy, while the last multiplier adds an additional factor of 1.5. Therefore, an oil droplet of the same size floats in water only 2 mm in 1 second and 12 cm in a minute. The record for slowness is held by a drop of water that "falls" in oil. With the same difference in density $\rho_0-\rho_1$, the viscosity of oil at a temperature of $20\ ^0 C$ is approximately 80 times higher than the viscosity of water. Taking into account the last multiplier, we will obtain a descent speed of the water droplet in oil of approximately 0.035\ mm per second, or 2 mm per minute, or, which is the same, 12\ cm\ per hour. The ascent speed of an air bubble will be 16.5 times higher than the ascent speed of an oil droplet of the same size and 900 times higher than the descent speed of a water droplet of the same size in oil, provided they are small, less than 0.5 mm.

We performed the calculation for a bubble with a diameter of 0.2\ mm. Smaller bubbles and droplets will move more slowly, with a decreasing speed proportional to the square of their size. Droplets twice as small will move 4 times slower, those 10 times smaller 100 times slower. An oil droplet with a diameter of 0.1\ mm will rise at a speed of 0.009\ mm/s, traveling 0.5\ mm in one minute and  3\ cm in an hour.

In light of these estimates, it becomes clear why, when using traditional technology, the suspension was left to settle for so long in the decanters. The waiting time depends on how slowly the water droplets move through the oil. If you don’t wait long enough, some water will stay mixed with the olive oil, which isn’t ideal. But if you wait too long, the oil can start to oxidize when exposed to air, which can affect its quality. It’s also worth noting that if tiny oil droplets remain in the water, they won’t affect the quality of the oil collected at the top of the decanter.

Today, the problem of slow droplets has been solved in a different way. The use of a centrifuge has made it possible to increase the effective value of g and, with it, the speed of ascent and descent of the droplets by thousands of times. However, the final separation of the water droplets requires further centrifugation in a special vertical centrifuge. It completes the oil extraction process, removing the smallest water droplets from it, and has no direct equivalent in the traditional process. However, it can be compared to the repeated decantation of the first-pressed oil for better separation from the water. 

Let's try to compare the different stages of olive oil production with the processes we are familiar with in coffee and wine production.

When talking about sparkling wines, such as champagne, it is believed that the bubbles should be as small as possible and, consequently, rise slowly. When decanting oil using the traditional method, it is preferable for the droplets to be larger, because they are more mobile. For this reason, an operation to encourage the merging of small droplets into larger ones was introduced into the olive oil extraction process. Modern technology can do without it, but this operation is still present, also as a sign of respect for the traditions of the craft. 

Physicists have studied the process of droplet fusion quite thoroughly. Its description was begun by Lord Rayleigh at the end of the 19th century and continued in the mid-20th century by the Soviet physicist Yakov Frenkel. He demonstrated \cite{Frenkel} that the fusion time of two identical droplets $\tau$ is proportional to the product of their radius $r$ and viscosity $\eta$ and inversely proportional to the surface tension coefficient $\sigma$:

\begin{equation}
 \tau \sim \frac{\eta r}{\sigma}.
\end{equation}                         
This formula is fundamental to powder metallurgy, but for understanding the oil extraction process, there's no need to go into such detail. In fact, the vertical centrifuge allows even the smallest droplets that haven't fused into larger drops to float. 

Let's return to the separation of oil and pomace in the horizontal centrifuge. In this case, these two components of the pulp move in opposite directions, passing through each other. We cannot describe this phenomenon as the movement of individual droplets in a medium, since the oil constitutes 12-15\% and the pomace represents 30\% to half of the total mass. Therefore, to some extent, we can talk about the flow of oil through the porous layer of the pomace, even if in a high-speed coordinate system connected to it. 

The process of liquid flow through a filter (porous medium) is described by the so-called Darcy's law. In the mid-19th century, the French engineer Henry Darcy  conducted the first experimental observations on the movement of water in pipes filled with sand.

These investigations gave rise to the theory of filtration, which is now successfully applied to describe the movement of liquids, gases, and their mixtures through solid bodies containing interconnected pores or fissures. In addition to creating the first modern water supply system in Europe in Dijon, Darcy formulated the so-called linear law of filtration \cite{Darcy}, which today bears his name. It applies to a homogeneous permeable medium -- is used in petroleum extraction, coffee brewing \cite{Varlamov}  and many other fields.  It connects, with a simple proportionality relationship, the volumetric flow rate of the fluid $Q$ with the pressure difference $ \Delta P$ between the boundaries of the filter (distance $L$) with area S filled by the porous medium:
\begin{equation}
	Q=\frac{\kappa S \Delta P}{\eta L}.   	
\end{equation}                   
Here, the coefficient $\eta$ characterizes the viscosity of the fluid, while the coefficient $\kappa$ is a characteristic only of the porous medium and is called the permeability coefficient (it has the dimension of area). 

Returning to oil filtration, formula (4) suggests that the speed at which it flows through the olive pomace is proportional to the pressure difference between the axis and the wall of the drum and that, thanks to the presence of centrifugal forces, it exceeds the pressure in a traditional press by hundreds of times. Therefore, this phase is also considerably accelerated by using a horizontal centrifuge. 

The skeptical reader might wonder what understanding physical processes contributes here. After all, olive oil was successfully pressed by farmers in antiquity and the Middle Ages, who had no knowledge of this science, nor of science in general. But it is precisely from this example that we see what science and the technology based on it have brought. 

The demand for olive oil has experienced strong growth over the past years. Annual production of extra-virgin olive oil in Italy is estimated at approximately 220,000 – 260,000 tons in recent years what means in market prices USD 2.5 billion. The global olive oil market size is ten times larger: it was USD 22.30 billion in 2022 and is anticipated to reach USD 33.12 billion by 2030. Evidently that it would be impossible to achieve such a scale of production using traditional old technologies.

Understanding the physics of processes used since antiquity has made it possible to invent a way to accelerate them with the help of a centrifuge and to combine previously separate processes into a single device. The result is a gain in terms of time, quality, and control of the olive oil production process. Work armed with knowledge is always more productive than work not supported by science and skill.

\section{Acknowledgements}
The authors express their deep gratitude to Lauro Morettini for introducing them to the Olive Oil Museum, which he founded, and for his continued interest in their work. They are also grateful to Michail Chernikov and Claudia Poggioni for their valuable comments. One of the authors (A.V.) wishes to thank Xiamen University, where this article was completed.

\end{document}